\documentclass[%
 reprint,
superscriptaddress,
 amsmath,amssymb,
 aps,prl,
]{revtex4-2}
\usepackage{hyperref}
\hypersetup{
    colorlinks=true,
    linkcolor=blue,
    filecolor=blue,      
    urlcolor=blue,
    citecolor=blue
    }
\usepackage{braket}
\usepackage{xcolor}
\usepackage{xfrac}
\usepackage{graphicx}
\usepackage{dcolumn}
\usepackage{dsfont}

\usepackage{bm}

\begin{document}

\preprint{APS/123-QED}

\title{Rydberg ion flywheel for quantum work storage}

\author{Wilson S. Martins}

\affiliation{%
Institut f\"ur Theoretische Physik, Universit\"at T\"ubingen, Auf der Morgenstelle 14, 72076 T\"ubingen, Germany
}%
\author{Federico Carollo}

\affiliation{%
Institut f\"ur Theoretische Physik, Universit\"at T\"ubingen, Auf der Morgenstelle 14, 72076 T\"ubingen, Germany
}%

\author{Weibin Li}%

\affiliation{School of Physics and Astronomy and Centre for the Mathematics and Theoretical Physics of Quantum Non-Equilibrium Systems, The University of Nottingham, Nottingham, NG7 2RD, United Kingdom}

\author{Kay Brandner}%

\affiliation{School of Physics and Astronomy and Centre for the Mathematics and Theoretical Physics of Quantum Non-Equilibrium Systems, The University of Nottingham, Nottingham, NG7 2RD, United Kingdom}

\author{Igor Lesanovsky}%

\affiliation{%
Institut f\"ur Theoretische Physik, Universit\"at T\"ubingen, Auf der Morgenstelle 14, 72076 T\"ubingen, Germany
}%
\affiliation{School of Physics and Astronomy and Centre for the Mathematics and Theoretical Physics of Quantum Non-Equilibrium Systems, The University of Nottingham, Nottingham, NG7 2RD, United Kingdom}

\date{\today}

\begin{abstract}
Trapped ions provide a platform for quantum technologies that offers long coherence times and high degrees of scalability and controllability. 
Here, we use this platform to develop a realistic model of a thermal device consisting of two laser-driven, strongly coupled Rydberg ions in a harmonic trap. 
We show that the translational degrees of freedom of this system can be utilized as a flywheel storing the work output that is generated by a cyclic thermodynamic process applied to its electronic degrees of freedom. 
Mimicking such a process through periodic variations of external control parameters, we use a mean-field approach underpinned by  numerical and analytical calculations to identify relevant physical processes and to determine the charging rate of the flywheel.  
Our work paves the way for the design of microscopic thermal machines based on Rydberg ions that can be equipped with both many-body working media and universal work storages. 
\end{abstract}

\maketitle

Developing new types of thermal machines that generate useful work at small length scales is a central topic in stochastic and quantum thermodynamics \cite{sei_2012, kos_2013, vin_and_2016, be_giu_2017, tru_mer_cre_2022, ar_2023}. 
Over the last decade, this area has seen remarkable progress driven by landmark experiments, in which  thermodynamic engine cycles were realized with microscopic objects such as single ions \cite{ro_daw_to_2016, po_sch_2019, hor_yum_2020}, nuclear spins \cite{pe_ba_2019}, nitrogen-vacancy centers in diamonds \cite{kla_be_2019} or large quasi-spin states of ultra-cold atoms \cite{bou_jens_2021}. 
Practical applications of such devices are, however, still limited, with two problems currently emerging as key challenges: first, scaling up the power of microscopic thermal machines without losing access to genuine features stemming from quantum effects \cite{mu_uma_2021, har_2015, uz_2016, nie_ger_2018, la_si_pe_2019, la_si_pe_2020,ta_fu_2021, ka_ha_ma_2022, so_man_ro_2022, ma_2022, kim_ji_2022, ko_dmy_sch_2023, ja_bea_cam_2016, li_bus_2018, chen_wa_2019, my_def_2020, fo_bus_2020, ke_fo_jin_2020, wa_ven_tal_2020, my_mc_def_2021, my_pe_ne_2022, eg_py_ke_ka_2022, ko_me_cues_2022}; second, identifying viable strategies to transfer the generated output to universal storage systems, which can be accessed by other devices \cite{zhan_ba_mey_2014, ro_daw_to_2016, po_sch_2019, hor_yum_2020, ca_bra_le_2020, she_che_2021, kim_ji_2022}. 

A promising approach to the first challenge is to replace working media with few degrees of freedom, such as single spins, with many-body systems, where collective effects can arise from the co-action of large numbers of constituents \cite{mu_uma_2021}. 
Recent theoretical and experimental studies have shown that the power of thermal devices can be significantly enhanced by exploiting, for example, many-body coherence in non-interacting systems, which can give rise to super-radiance and related phenomena \cite{har_2015, uz_2016, nie_ger_2018, la_si_pe_2019, la_si_pe_2020,ta_fu_2021, ka_ha_ma_2022, so_man_ro_2022, ma_2022, kim_ji_2022, ko_dmy_sch_2023}, or interactions and quantum many-body statistics in ultra-cold atomic systems \cite{ja_bea_cam_2016, li_bus_2018, chen_wa_2019, my_def_2020, fo_bus_2020, ke_fo_jin_2020, wa_ven_tal_2020, my_mc_def_2021, my_pe_ne_2022, eg_py_ke_ka_2022, ko_me_cues_2022}. 
Strongly interacting Rydberg atoms and ions provide another, yet relatively unexplored, platform to implement quantum thermal machines with many degrees of freedom \cite{car_gam_bran_2020}. 
These systems show a rich phenomenology and can be realized in experiments with a high degree of control and access to internal state variables \cite{hi_li_fa_2017, hi_po_fa_2017, mok_hei_2020}.
Rydberg ions, in particular, offer state-dependent interaction together with long-time stability \cite{li_le_2012, gam_le_li_2019, gam_li_sch_2020,  gam_zhan_chi_2020}.

In addition, interactions among Rydberg states give rise to an accurately controllable coupling between translational and internal electronic degrees of freedom \cite{gam_zha_hen_2021,ma_jo_le_2023}. 
This feature can be exploited to approach the second challenge in the development of practically applicable quantum thermal machines.
Inspired by earlier experiments with single-body systems \cite{ro_daw_to_2016, po_sch_2019}, the key idea here is to perform an engine cycle with the electronic subsystem, while the external degrees of freedom act as a storage for mechanical work akin to the flywheel of a macroscopic engine. 
In this article, we take a first step towards exploring this idea. 
Using a minimal model consisting of two harmonically trapped Rydberg ions, whose realization was recently reported in Ref. \cite{zhang_po_wei_2020}, we show that usable work in the form of electronic excitations can be transferred to a vibrational degree of freedom, which forms our flywheel. 
In lieu of a thermodynamic engine cycle, our device is driven by periodic modulations of the dynamical parameters that control the effective Hamiltonian of the electronic working medium. 
Our central aim is to demonstrate that Rydberg ion systems, under realistic conditions, provide a potent platform for thermal devices that have access to quantum many-body effects and, at the same time, are capable of delivering significant output to externally accessible work storages.  \\ 

\newcommand{\xrel}{x_\mathrm{rel}}
\newcommand{\orel}{\omega_\mathrm{rel}}
\newcommand{\ph}{\mathrm{ph}}
\newcommand{\el}{\mathrm{el}}

\begin{figure*}
\centering
\includegraphics[scale = 0.52]{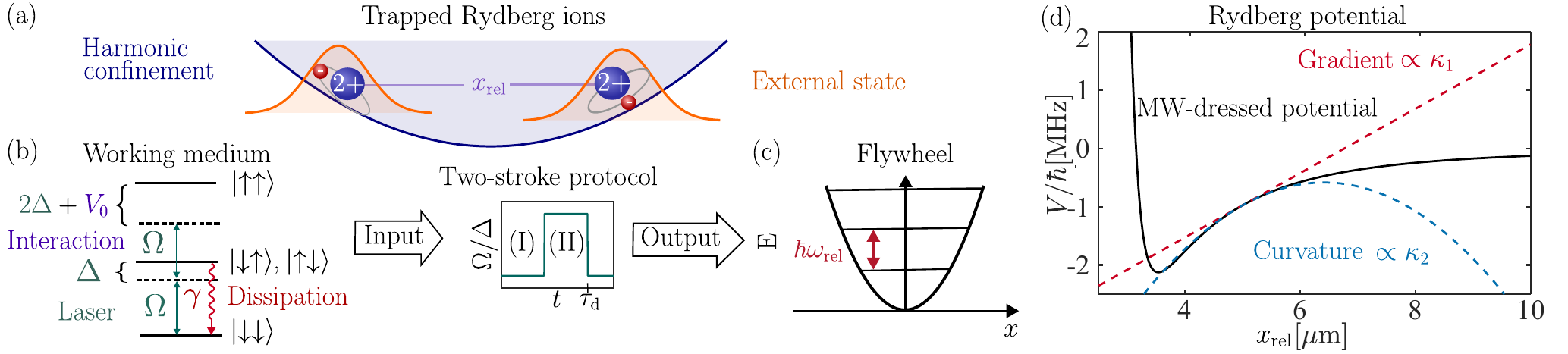}
\vspace{-0.75cm}
\caption{\textbf{Rydberg ion flywheel.} (a) Setup. Two ions with a doubly charged core and one bound Rydberg electron are confined in a harmonic trap that gives rise to a typical separation of $x_\mathrm{rel}\simeq 5\mu\mathrm{m}$. 
(b) Electronic degrees of freedom (working medium). Each ion is modelled by two electronic states (ground state $\Ket{\downarrow}$, Rydberg state $\Ket{\uparrow}$). The figure shows the energy level scheme and transitions associated with the electronic states of two Rydberg ions. 
A laser drives the transition between ground and Rydberg states with Rabi frequency $\Omega$ and detuning $\Delta$; spontaneous decay of the Rydberg states occurs with rate $\gamma$. 
If both ions are excited, the electrostatic interaction among Rydberg states, $V_0$, shifts the energy of the doubly excited state $\Ket{\uparrow\uparrow}$ 
\cite{lu_flei_co_2001, ur_john_2009, co_pi_2010}, cf. Eq. \eqref{int_ham_exp}. 
Cyclically changing $\Omega$ or $\Delta$ (period $\tau_\mathrm{d}$) makes it possible to modulate the population of the double excited state periodically and thus the force between the ions. 
This process excites the vibrational degree of freedom and thereby charges the flywheel. 
For concreteness, we consider a two-stroke protocol with period $\tau_\mathrm{d}$, where the control parameter switches between two fixed values. 
(c) Vibrational degree of freedom (flywheel). 
At low energies $E$, the vibrational mode, where the output of the working medium is stored, can be described as a harmonic oscillator with frequency $\omega_\mathrm{rel}$.
(d) Two-ion microwave (MW) dressed potential $V_\mathrm{int}(x_\mathrm{rel})$ as a function of the distance $\xrel$. 
Around the ions' equilibrium positions the potential is characterized by its gradient and curvature, which are proportional to the parameters $\kappa_1$ and $\kappa_2$.
The figure is based on the setup of Ref.~\cite{gam_zhan_chi_2020}, using dressed Rydberg states of ${}^{88}\mathrm{Sr}^+$.}
\label{model}
\end{figure*}

\noindent
\textbf{Model.}-- We consider the setup of Fig. \ref{model}.  
Two Rydberg ions with mass $m$ and charge $e$ are trapped in an isotropic harmonic potential with strength $\omega$. 
We focus on the longitudinal motion of the ions along their connecting axis, which is governed by the potential $V_\mathrm{ions}(x_1, x_2) = \frac{1}{2}m\omega^{2}(x_1^2 + x_2^2) + V_\mathrm{el}(\xrel)$.
Here, $x_k$ with $k=1,2$ are the positions of the ions, $\xrel = |x_1-x_2|$ and $V_\mathrm{el}(\xrel) = e^2/4\pi \epsilon_0 \xrel$ denotes the electrostatic potential, where $\epsilon_0$ is the vacuum permittivity \cite{mu_li_le_2008, sch_fel_kol_2011}. 
At low energies, the ions oscillate around their equilibrium positions $x^0_k$. 
The potential $V_{{{\rm ions}}}(x_1,x_2)$ can then be expanded to second order in the displacements $\delta x_k = x_k - x^0_k$ \cite{la_ba_ra_2016, en_ber_kees_2016, lie_de_ba_2018, gam_zha_hen_2021}. 
After separating the center of mass and relative motion and quantizing the relative displacement by making the replacement 
\begin{equation}
\delta x_\mathrm{rel}/\ell_0 = (\delta x_1-\delta x_2)/\ell_0 \rightarrow x = (a^\dagger + a)/\sqrt{2},
\end{equation}
we obtain the Hamiltonian $H_\mathrm{ph} = \hbar \omega_\mathrm{rel} \left(a^\dagger a + 1/2 \right) \label{free_ham_1}$
for the vibrational dynamics. 
Here, $\omega_\mathrm{rel}= \sqrt{3}\omega$ is the reduced frequency, $\ell_0 = \sqrt{2\hbar/m\omega_\mathrm{rel}}$ denotes the characteristic length scale of the oscillator and $a$ and $a^\dagger$ are the usual annihilation and creation operators; for details, see Ref.~\cite{supp}.  
The internal degrees of freedom of the ions are modeled as two-level systems with excited Rydberg state $\Ket{\uparrow}$ and ground state $\Ket{\downarrow}$ \cite{low_2012, mar_min_2017}.
The transition between these states is driven by a laser with Rabi frequency $\Omega$ and detuning $\Delta$. 
In the rotating frame of the laser, the free electronic dynamics are described by the Hamiltonian $H_\mathrm{el} = \hbar  \sum_{k=1}^2 (\Delta n_k + \Omega \sigma_k^x)$, where $\sigma^x_k = \Ket{\uparrow_k}\Bra{\downarrow_k} + \Ket{\downarrow_k}\Bra{\uparrow_k}$ and $n_{k} = \Ket{\uparrow_k}\Bra{\uparrow_k}$. 

When excited to Rydberg states, the ions are subject to the interaction $H_\mathrm{int} = V_\mathrm{int}(x_\mathrm{rel})n_1 n_2$ \cite{ga_1988, low_2012, bro_ba_2016}. 
This interaction represents a correction to the potential $V_\mathrm{ions}$, since its magnitude is small compared to the electrostatic repulsion \cite{saf_wal_mo_2010}.
This Hamiltonian leads to a shift of the effective energy levels of the ions, see Fig. \ref{model}(b), and a state-dependent force that couples their external and internal degrees of freedom. The interaction potential between Rydberg ions is typically of dipolar or van-der-Waals type \cite{wal_saf_2008, mu_li_le_2008,  saf_wal_mo_2010, gam_le_li_2019}. 
This interaction is, however, generally weaker than that between neutral Rydberg atoms due to a scaling of the electric dipole with the inverse nuclear charge, $Z^{-1}=1/2$. Strong interactions can nevertheless be realized through microwave (MW) dressing. The gradient and curvature of the resulting potentials can be accurately controlled in experiments \cite{mu_li_le_2008, vo_li_mokh_2019, zhang_po_wei_2020}, see Fig. \ref{model}(d). 
Upon expanding the such a potential to second order in the relative displacement $\delta x_\mathrm{rel}$ and quantizing the vibrational degree of freedom as before, we obtain the effective interaction Hamiltonian 
 \begin{equation}
    H_\mathrm{int} = \left( V_0 + \hbar \kappa_1 x + \hbar \kappa_2 x^2 \right) n_1 n_2 = \hbar W(x) n_1 n_2.
    \label{int_ham_exp}
 \end{equation}
In this Hamiltonian, $V_0 = V_\mathrm{int}(x_\mathrm{rel}^0)$ sets the baseline for the interaction strength, with $x_\mathrm{rel}^0=|x^0_1-x^0_2|$ being the equilibrium distance between the ions; $\hbar\kappa_1 = \ell_0 V'_\mathrm{int}(x_\mathrm{rel}^0)$ and $\hbar\kappa_2 = \ell_0^2 V''_\mathrm{int}(x_\mathrm{rel}^0)/2$ are proportional to the gradient and the curvature of the potential. 

To account for the spontaneous decay of excited Rydberg ions, we complete our model by including a Lindblad-type dissipation super-operator with the form $\mathcal{L}[\bullet] = \gamma \sum_{k = 1}^2 (\sigma^-_k\bullet \sigma^+_k - \frac{1}{2}\{n_k, \bullet\})$, where $\sigma_k^- = \Ket{\downarrow_k}\Bra{\uparrow_k}$ and $\sigma^+_k = \Ket{\uparrow_k}\Bra{\downarrow_k}$ are local jump operators, curly brackets denote the anti-commutator and $\gamma$ is a decay rate, see Fig.\ref{model}(b). 
Hence, the state $\varrho$ of the system follows the quantum master equation 
\begin{equation}\label{ME}
\dot{\varrho} = -\frac{i}{\hbar}[H, \varrho] + \mathcal{L}[\varrho]
\end{equation}
with the full Hamiltonian $H = H_\mathrm{ph} + H_\mathrm{el} + H_\mathrm{int}$.

\noindent
\textbf{Mean-field dynamics.}--
To explore the dynamics of our model, we proceed with a mean-field approximation, where correlations between internal and external degrees are neglected. 
The validity of such an approximation is discussed in the Supplemental Material \cite{supp}, where we  compare mean-field results with numerical ones obtained by truncating the Fock space. We assume that $\varrho = \varrho_\ph\otimes\varrho_\el$, where $\varrho_\ph$ and $\varrho_\el$ describe the vibrational and electronic dynamics, respectively.  
These states follow the mean-field equations 
\begin{equation}
    \dot{\varrho}_\ph = -\frac{i}{\hbar}[\tilde{H}_\ph,\varrho_\ph],
	\quad
    \dot{\varrho}_\el = -\frac{i}{\hbar}[\tilde{H}_\el,\varrho_\el] + \mathcal{L}[\varrho_\el]
    \label{uni_ev}
\end{equation}
with $\tilde{H}_\ph = H_\ph + \hbar W(x)s_{nn}$, $\tilde{H}_\el = H_\el + \hbar n_1n_2 \langle W(x) \rangle$. 
Here, the operator $W(x)$ was defined in Eq.~\eqref{int_ham_exp} and the variable $s_{nn} = \left \langle n_1 n_2 \right \rangle$, which corresponds to the population of the double excited state $\ket{\uparrow\uparrow}$, encodes the driven-dissipative dynamics of the electronic sub-system, see Fig.~\ref{model}(b). 
Throughout this article, we use the short-hand notation $\langle\bullet\rangle = \mathrm{Tr}[\bullet\varrho_\ph\otimes\varrho_\el]$. 

In the mean-field picture, the average relative displacement follows the equation of motion 
\begin{equation}
    \langle \ddot{x} \rangle + \orel(\orel+ \kappa_2 s_{nn})\left \langle x \right \rangle = -\orel\kappa_1 s_{nn} \, .
    \label{imp_eq}
\end{equation}
This resembles the equation of motion for a driven harmonic oscillator. However, 
we note that here the variable $s_{nn}$ is time-dependent and even depends on the position $\langle x\rangle$ itself; for the complete set of mean-field equations, see Ref.~\cite{supp}.
This observation suggests that, in order to inject maximal power into the flywheel, the driving protocol for the working medium should be chosen such that $s_{nn}$ oscillates with the eigenfrequency $\omega_\ph= \sqrt{\orel(\orel+\kappa_2 s_{nn})}$ of the oscillator \eqref{imp_eq}. 
To meet this condition, we focus on the regime, where $\kappa_2/\orel\ll 1$ and $\omega_\ph\approx \orel$ becomes nearly independent of the electronic state variable $s_{nn}$. 
This situation can be realized by tuning the MW-dressed potential and the trap strength $\omega$ so that the equilibrium distance between the ions $\xrel^0$ comes close to the distance $x^\ast_\mathrm{rel}$, where the curvature of the interaction potential vanishes \cite{gam_zhan_chi_2020, gam_li_sch_2020}. 
For the parameters used in Fig.~\ref{model}(d), we have $x^\ast_\mathrm{rel}\simeq 3.94\;\mu\mathrm{m}$ for $\omega \simeq 2\pi \times 145\;\mathrm{kHz}$,  which is a realistic value in typical experiments with Rydberg ions \cite{hi_li_fa_2017, hi_po_fa_2017, zhang_po_wei_2020}.
The baseline interaction strength, the gradient of the interaction potential, the Rydberg state decay rate, and the characteristic length of the vibrational sub-system then become $V_0/\hbar \simeq - 1.8\;\mathrm{MHz}$, $\kappa_1 = \gamma \simeq 0.1\;\mathrm{MHz}$ and $\ell_0 \simeq 0.1\;\mu\mathrm{m}$. 
In the following analysis, we use these values as a reference. 

\newcommand{\td}{\tau_\mathrm{d}} 
\newcommand{\lc}{\mathrm{lc}}
\newcommand{\mx}{\langle x\rangle}

\begin{figure}
\centering
\includegraphics[scale = 0.4]{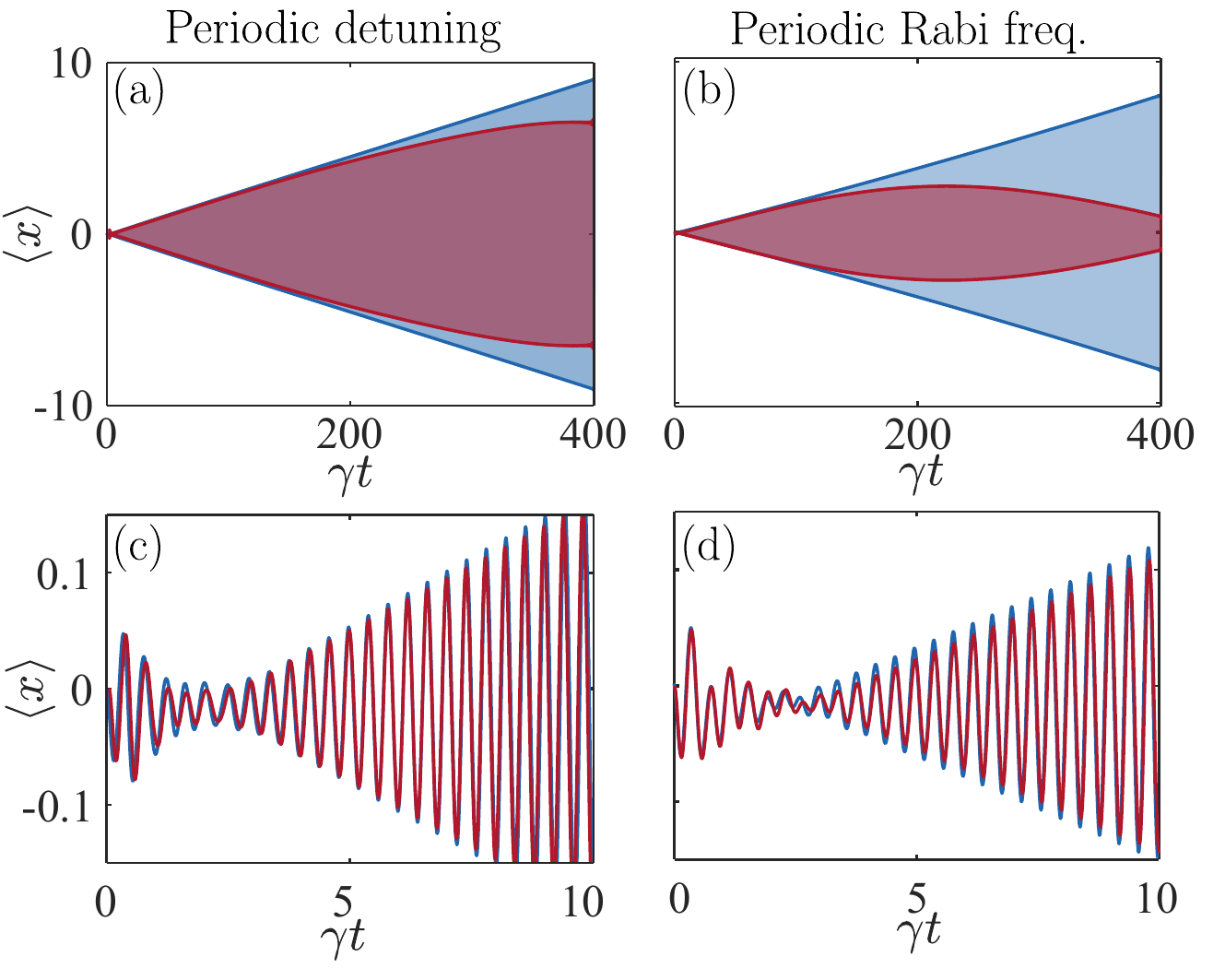}
\vspace{-0.75cm}
\caption{\textbf{Charging the flywheel.} 
(a) Average distance between the Rydberg ions in units of $\ell_0$ as a function of time for two different values of the curvature parameter $\kappa_2$; 
the detuning $\Delta$ switches cyclically between $\Delta_\mathrm{min} =0$ and $\Delta_\mathrm{max}= 9\gamma$, while the Rabi frequency $\Omega = 8\gamma$ is fixed. 
(b) Same plot as in (a) with $\Omega$ switching between $\Omega_\mathrm{min}=2\gamma$ and $\Omega_\mathrm{max}=8\gamma$ and $\Delta= 9\gamma$ fixed.  
The blue curves were obtained with $V_0 = -18\hbar\gamma$ and $\kappa_2 = 0$; the red ones with $V_0 = -17\hbar\gamma$ and $\kappa_2 = -0.1\gamma$. 
Figures (c) and (d) show the oscillatory short-time dynamics of the flywheel. 
For all plots, we have chosen $\omega=9\gamma$, $\kappa_1=\gamma$ and initially set $\varrho_\ph = \Ket{0}\Bra{0}$ and $\varrho_\el = \Ket{\downarrow\downarrow}\Bra{\downarrow\downarrow}$, where $\Ket{0}$ is the ground state of the oscillator.}
\label{pos_op}
\end{figure}

\noindent  
\textbf{Results.}-- To charge our flywheel, we vary either the Rabi frequency $\Omega$ or the detuning $\Delta$ of the laser according to the periodic two-stroke protocol shown in Fig.~\ref{model}(b); the switching occurs at $t=\td/2$ and the period is set to $\td = 2\pi/\orel$; the detuning can be controlled, for instance, through external fields affecting the energy levels of the ions. 
The level scheme in Fig.~\ref{model}(b) shows that the alignment between the effective energy levels of the working medium and the transitions driven by the laser depends on both $\Omega$ and $\Delta$. 
As a result, both parameters affect the rate at which excitations are created in the electronic system and can therefore be used to imprint a periodic modulation on the double-excitation probability $s_{nn}$, which controls the repulsion force between the ions. 
This mechanism enables a continuous transfer of energy from the working medium to the flywheel, which leads to the gradual increase of its oscillation amplitude seen in Fig.~\ref{pos_op}.
In line with our physical picture, the charging process is suppressed, and even reversed, at long times for $\kappa_2\neq 0$, as the flywheel is shifted out of resonance with the driving.
We note that the runtime of the flywheel is limited since the width of the ion wave-packages must be much smaller than their equilibrium distance to ensure the validity of the approximated interaction potential. 
For the parameters used in Fig.~\ref{pos_op}, this condition is met for $t\leq 400/\gamma$. 
In experiments also the phononic modes experience some dissipation. However, as we show in the Supplemental Material \cite{supp}, parameter regimes can be identified in which this is negligible.

The qualitative behavior of the mean distance $\langle x \rangle$ can be further understood from the high-frequency limit. 
To this end, we first observe that the mean-field equations \eqref{uni_ev} decouple for $\kappa_1=\kappa_2=0$.
After some relaxation time $\tau_0$, which is essentially determined by $\gamma$, the electronic state $\varrho_\el$ then settles to a unique limit cycle, which satisfies $\varrho_\el^\lc(t) = \varrho_\el^\lc(t+\td)$ \footnote{This statement follows from the observation that the Lindblad generator $\mathcal{L}$ in Eq.~\eqref{uni_ev} has only one eigenvalue with vanishing real part as can be easily confirmed by inspection; for further details see \cite{men_bran_2019}.}.
Hence, $s_{nn}$ acquires the same periodicity as the driving. 
For $0<|\kappa_1|\ll\orel$, the oscillator \eqref{imp_eq} is affected by driving only over a large number of periods. 
Thus, $\mx$ remains $\td$-periodic on short time scales and develops modulations on some longer scale   $\tau_\mathrm{mod}=1/\epsilon$. 
That is, $\mx = \mx(\epsilon t, \orel t) $ can be written as a Fourier series with slowly drifting coefficients, 
\begin{equation}\label{eq:xansatz1}
    \mx =\sum\nolimits_{n\in\mathbb{Z}} c_n(\epsilon t)
    e^{i n\orel  t}
    = \epsilon t \cdot \mx_1(\orel t) + \mathcal{O}(\epsilon^2).
\end{equation}
Here, we have expanded in $\epsilon$ and used that $\mx = \langle \dot{x}\rangle =0$ at $t=0$ so that $\mx\rightarrow 0$ for $\epsilon\rightarrow 0$; note that we understand any function of $\orel t$ to be $2\pi$-periodic in this argument.  
The time-dependence of $\mx$ now carries over to the mean-field Hamiltonian $\tilde{H}_\el$, which is thus no longer strictly periodic. 
However, if $\epsilon\tau_0\ll 1$, the working medium still follows its instantaneous limit-cycle on the long time scale. 
Thus, we have $\varrho_\el \approx \varrho_\el^\lc(\epsilon t, \orel t)$ and 
$s_{nn} \approx s_{nn}(\epsilon t, \orel t) = \frac{1}{2}\sum_{n\in\mathbb{Z}} d_n(\epsilon t)e^{in\orel t}$
for $t> \tau_0$. 
The scale of the modulation rate $\epsilon$ can now be determined self-consistently.
Inserting the ansatz $\mx = \mx(\epsilon t, \orel t)$ into Eq.~\eqref{imp_eq} and setting $\kappa_2=0$ gives 
\begin{align}
\mx & = -\kappa_1 \int_0^t \!\! dt' |d_1(\epsilon t')|\cos[\orel t + \varphi_1(\epsilon t')] + \mathcal{O}(\kappa_1/\orel)\nonumber\\
    & = -\kappa_1 |d_1(0)| t \cdot \cos[\orel t + \varphi_1(0)] + \mathcal{O}(\kappa_1/\orel, \epsilon), 
    \nonumber
\end{align}
where $\varphi_1$ is a phase shift. 
Comparing this results with Eq.~\eqref{eq:xansatz1} shows that $\epsilon$ must be of the same order of magnitude as $|\kappa_1 d_1(0)|$.
Finally, for $\gamma\ll\orel$, the electronic system is barely able to follow the driving protocol and the unperturbed oscillation amplitude $d_1(0)$ of $s_{nn}$ is of order $\gamma/\orel$. 
We then have $\epsilon\tau_0 \sim \epsilon/\gamma \sim |\kappa_1|/\orel \ll 1$, which shows that our estimate is self-consistent in the high-frequency regime. 

The above argument still holds for $\kappa_2\neq 0$ as long as $|\kappa_2|/\orel \ll 1$. 
Quite intuitively, it shows that the charging rate $\epsilon$ of our flywheel is essentially determined by the strength of its interaction with the working medium and frequency of the external harmonic trap, $\omega = \orel/\sqrt{3}$. 
However, the specific choice of the driving protocol does not play a dominant role. 
We note that the estimate $\epsilon\sim\gamma|\kappa_1|/\orel$ is also in good agreement with our numerical simulations; for the parameters chosen in Fig.~\ref{pos_op}, the charging rate is $\epsilon\approx \gamma/40$, while $\gamma|\kappa_1|/\orel \approx \gamma/16$.

\begin{figure}
    \centering
    \includegraphics[scale = 0.42]{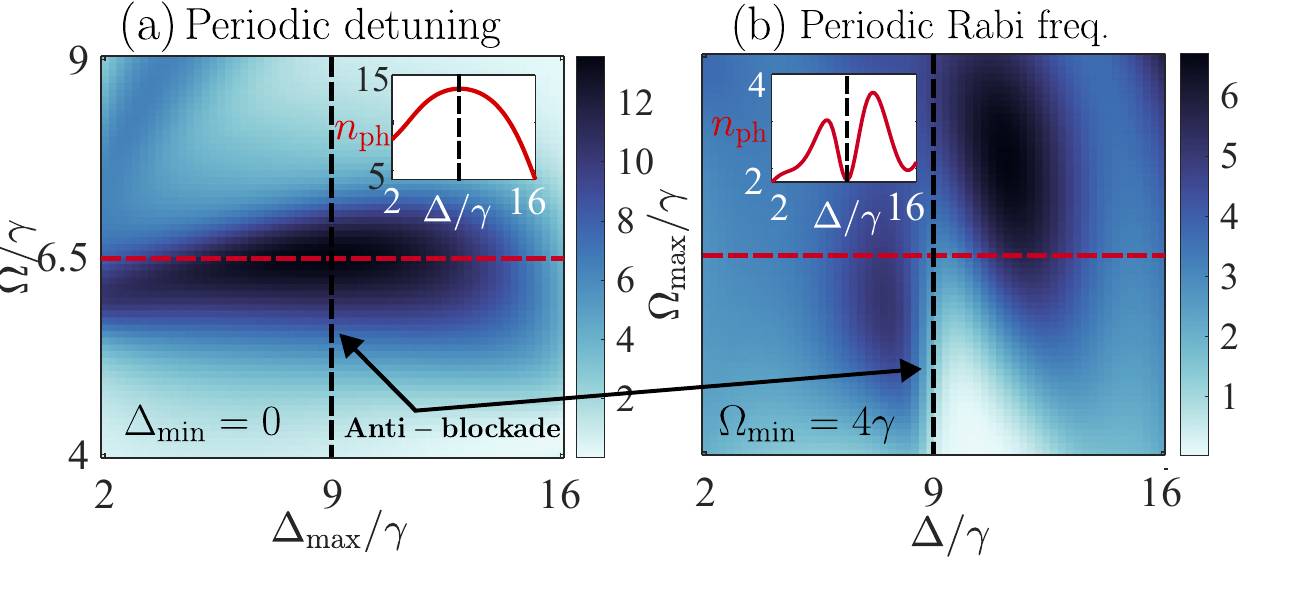}
    \vspace{-0.75cm}
    \caption{\textbf{Energy of the flywheel.}
The plots show the average excitation number $n_{{{\rm ph}}}$ after a runtime of  $t = 100/\gamma$ for both driving modes as a function of (a) the Rabi frequency and the maximal detuning and (b) the maximal Rabi frequency and the fixed detuning. 
The insets show cuts through the density plots along the horizontal dashed lines. 
The vertical dashed lines indicate the anti-blockade condition, see main text for details.     
For all plots, we have chosen the parameters $\omega = 9\gamma$,  $V_0 = -18\hbar\gamma$, $\kappa_1 = \gamma$ and $\kappa_2 = 0$.}
    \label{excit}
\end{figure}

To further explore the phenomenology of our model, we now analyze the energy content of our flywheel, which is proportional to the mean excitation number 
$n_\mathrm{ph} = \left \langle a^\dagger a \right \rangle$. 
This quantity is plotted in Fig.~\ref{excit} as a function of the laser parameters for both driving modes and a runtime of $t=100/\gamma$. 
The main features of these plots can be understood as follows. 

For periodically changing detuning, we observe a pronounced maximum when $\Delta_{{{\rm max}}}$ meets the so-called anti-blockade condition $2\Delta + V_0/\hbar = 0$ \cite{at_ph_pa_2007, am_tho_2010, youn_bou_2018, fes_lo_niko_2022, ma_ma_le_2022}. 
The transition between ground and double excited state of the working medium is then resonant with the laser during the second stroke of the protocol, which leads to a strong increase of $s_{nn}$, see Fig.~\ref{model}(c). 
Leaving the anti-blockade regime in the first stroke by changing the detuning so that $\Delta\ll \Delta_{{{\rm max}}}$ leads to a sharp decay in the double excitation probability due to spontaneous decay. 
As a result, $s_{nn}$ develops a large oscillation amplitude, which gives rise to a large charging rate. 

By contrast, if the system is driven through the Rabi frequency of the laser, the anti-blockade regime features a dip in the energy of the flywheel. 
This observation can be explained by considering the three relevant eigenstates, $\Ket{\downarrow\downarrow}$, $\Ket{\uparrow\uparrow}$ and $\Ket{S}\propto \Ket{\downarrow\uparrow}+\Ket{\uparrow\downarrow}$, of the reduced mean-field Hamiltonian $\tilde{H}^0_{{{\rm el}}}=  \hbar\Delta\sum_{k=1}^2 n_k + V_0 n_1 n_2$; the anti-symmetric superposition of the single excited states does not couple to the laser due to the permutation symmetry of $\tilde{H}_{{{\rm el}}}$. 
If the anti-blockade condition is met, the state $\Ket{\downarrow\downarrow}$ and the state $\Ket{\uparrow\uparrow}$ are both ground states of the Hamiltonian $\tilde{H}^0_{{{\rm el}}}$.
The superpositions $\Ket{D} \propto \Ket{\downarrow\downarrow} - \Ket{\uparrow\uparrow}$ and $\Ket{B} \propto \Ket{\downarrow\downarrow} + \Ket{\uparrow\uparrow}$ then correspond to a dark and a bright state of the system, respectively \cite{rao_mol_2013}. 
Since neither of these states depends on $\Omega$, the dark state becomes a stable fixed point of the dynamics, in which the working medium is effectively trapped; note that such a fixed point does not exist if the anti-blockade condition is periodically lifted by changing the detuning and the emergence of this dark-state has no classical counterpart. 
This mechanism suppresses the oscillation amplitude of $s_{nn}$, and thus the charging rate.
We note that the above argument, though covering the dominant physical process, does not account for spontaneous decay or the modulation of the electronic Hamiltonian through the position of the flywheel. 
Therefore, we still expect the charging rate to remain finite if the anti-blockade condition is met, as our simulations show. 

\noindent
\textbf{Concluding perspectives.}--
In this work, we have analyzed a minimal yet realistic model of an integrated thermal machine based on laser-driven Rydberg ions. 
The electronic degrees of freedom of the ions provide a working medium for a thermodynamic cycle, here mimicked through periodic variations of external control parameters. Their translational degrees of freedom, on the other hand, act as a flywheel storing the generated work output. 
To what extent this output is accessible to secondary devices will depend on the specifics of the coupling mechanism. 
If arbitrary unitary transformations can be applied to extract work from the flywheel, the maximal accessible energy is given by its \emph{ergotropy} \cite{alla_ba_nieu_2004}, which, in the mean-field regime, is, up to a constant shift \cite{erg}, equivalent to the internal energy plotted in Fig.~\ref{excit}.

Our study demonstrates that Rydberg-ion systems are a viable experimental platform for microscopic thermal devices that feature genuine quantum effects and are capable of delivering output to an external storage system. 
Furthermore, our model can, in principle, be scaled up to a many-body device by replacing the pair of ions with an ionic Wigner crystal \cite{ja_1998, po_ci_2004, den_po_ci_2005}, where selected phonon modes play the role of the flywheel. 
This step, which promises to reveal a rich phenomenology arising from many-body effects, along with a complete thermodynamic analysis of our model and the integration of proper thermodynamic cycles driven by thermal rather than coherent energy sources rather are left to future research. 
Our results here provide both a well-defined starting point and a valuable benchmark for these investigations.

\begin{acknowledgments}
\textbf{Acknowledgments:}
We acknowledge funding from the Deutsche Forschungsgemeinschaft (DFG, German Research Foundation) through the Research Unit FOR 5413/1, Grant No. 465199066. This project has also received funding from the European Union’s Horizon Europe research and innovation program under Grant Agreement No. 101046968 (BRISQ). F.C.~is indebted to the Baden-W\"urttemberg Stiftung for the financial support by the Eliteprogramme for Postdocs. This work was supported by the University of Nottingham and the University of T\"{u}bingen’s funding as part of the Excellence Strategy of the German Federal and State Governments, in close collaboration with the University of Nottingham. 
KB acknowledges support from the University of Nottingham through a Nottingham
Research Fellowship. This work was supported by the Medical Research Council [grant
number MR/S034714/1]; and the Engineering and Physical Sciences Research Council
[grant numbers EP/V031201/1 and EP/W015641/1].
\end{acknowledgments}

\bibliography{ryd_ion_enginev2}

\begin{thebibliography}{78}
\bibitem [{erg()}]{erg}%
  \BibitemOpen
  \href@noop {} {}\bibinfo {note} {The ergotropy is defined by $\mathcal{E} =
  \mathrm{tr}[\varrho_t H] - \min_U \mathrm{tr}[H U\varrho U^\dagger]$, with
  $U^\dagger U = \mathds{1}$. In the present system, for an initial phononic
  state $\Ket{0}\Bra{0}$ it reduces to $\mathcal{E} = \mathrm{tr}[H_\ph
  \varrho_\ph] - \Bra{0}H_\ph\Ket{0} = \mathrm{tr}[H_\ph \varrho_\ph] - \hbar
  \orel/2$}\BibitemShut {NoStop}%
\end{thebibliography}

\end{document}


\preprint{APS/123-QED}

\renewcommand\thesection{S\arabic{section}}
\renewcommand\theequation{S\arabic{equation}}
\renewcommand\thefigure{S\arabic{figure}}
\setcounter{equation}{0}
\setcounter{figure}{0}

\onecolumngrid

\newpage

\setcounter{page}{1}
\widetext
\begin{center}
{\Large SUPPLEMENTAL MATERIAL}
\end{center}
\begin{center}
\vspace{0.8cm}
{\Large {Rydberg ion flywheel for quantum work storage}}
\end{center}

\begin{center}
Wilson S. Martins$^1$, Federico Carollo$^1$, Weibin Li$^2$, Kay Brandner$^2$ and Igor Lesanovsky$^{1,2}$ 
\end{center}
\begin{center}
$^1${\em Institut f\"{u}r Theoretische Physik,  Universit\"{a}t T\"{u}bingen, Auf der Morgenstelle 14, 72076 T\"{u}bingen, Germany,}\\
{\em Auf der Morgenstelle 14, 72076 T\"ubingen, Germany}\\
$^2${\em School of Physics and Astronomy and Centre for the Mathematics}\\
{\em and Theoretical Physics of Quantum Non-Equilibrium Systems,}\\
{\em  The University of Nottingham, Nottingham, NG7 2RD, United Kingdom}\\

\end{center}

In this supplemental material, we provide further details on the approximation that were made in the main text to describe our flywheel. 
We further discuss the range of validity of these approximations with the help of numerically exact calculations.

\section{ \label{exp_pot} Effective potential for external degrees of freedom}

\newcommand{\xcm}{x_{{{\rm cm}}}}
\newcommand{\cm}{{{{\rm cm}}}}
\newcommand{\ext}{{{{\rm ext}}}}
\newcommand{\rel}{{{{\rm rel}}}}
\newcommand{\av}[1]{\langle #1\rangle}
\newcommand{\xrel}{x_\mathrm{rel}}
\newcommand{\orel}{\omega_\mathrm{rel}}
\newcommand{\ph}{\mathrm{ph}}
\newcommand{\el}{\mathrm{el}}

As described in the main text, we consider two identical Rydberg ions with mass $m$ and charge $e$. 
The ions are confined in a trapping potential that is tuned to be isotropic and nearly harmonic with frequency $\omega$ \cite{hi_po_zhan_2019}. 
Upon neglecting charge-dipole interactions, which can in principle be treated via a second-order perturbation theory \cite{vo_li_mokh_2019}, and focusing on the longitudinal motion of the ions, we arrive at the net potential 
\begin{equation}\label{SPot}
    V_\mathrm{ions}(x_1, x_2) = \frac{1}{2}m\omega^{2}(x_1^2 + x_2^2) + V_\mathrm{el}(\xrel),
\end{equation}
where $x_k$ with $k=1,2$ are the positions of the ions along their connecting axis and $V_{{{\rm el}}}=e^2/4\pi\epsilon_0 \xrel$ describes the electrostatic interaction between them; $\xrel=|x_1-x_2|$ denotes the distance between the ions and $\epsilon_0$ the vacuum permittivity. 
By introducing the center-of-mass coordinate $\xcm= (x_1+x_2)/2$, the Hamiltonian of the external degrees of freedom can be split into two contributions 
\begin{equation}\label{SPot}
    H_\ext = H_\cm + H_\ph \quad\text{with}\quad
    H_\cm = \frac{p^2_\cm}{2M} +\frac{1}{2}M\omega\xcm^2, \quad
    H_\ph = \frac{p_\rel^2}{2\mu} + V_\rel(\xrel), \quad
    V_\rel(\xrel) = \frac{1}{2}\mu\omega \xrel^2 + V_{{{\rm el}}}(\xrel).  
\end{equation}
Here, $p_\cm$ and $p_\rel$ are the canonical momenta conjugate to $\xcm$ and $\xrel$ and $M=2m$ and $\mu=m/2$ denote the total and the reduced mass. 

We now focus on the vibrational motion of the ions, which is described by the Hamiltonian $H_\ph$. 
The potential $V_\rel$ has a stable minimum at $\xrel^0  = (e^2/2 \pi m \omega^{2}\epsilon_{0})^{1/3}$. 
Hence, for sufficiently low energies, $V_\rel$ can be expended to second order in the displacement $\delta\xrel = \xrel - \xrel^0$ \cite{ja_1998}. 
After neglecting trivial offsets and quantizing the relative coordinate by means of the replacements 
\begin{equation}
\delta\xrel \rightarrow \ell_0(a^\dagger + a)/\sqrt{2}
    \quad\text{and}\quad 
p_\rel\rightarrow \wp_0 (a^\dagger- a)/\sqrt{2},
\end{equation}
where $\ell_0 =\sqrt{\hbar/2\mu\orel}$ and $\wp_0=\sqrt{\hbar\mu\orel/2}$ denote the characteristic length and momentum scales of the system, we obtain the reduced Hamiltonian 
\begin{equation}
H_\ph = \hbar\orel (a^\dagger a + 1/2),
\end{equation}
where $a^\dagger$ and $a$ are the usual Bosonic creation and annihilation operators and $\orel = \sqrt{3}\omega$ denotes the frequency of the vibrations.

\section{\label{int_cor} Mean-field approach and comparison with completely computational solutions}

Within the mean-field approximation, the vibrational and electronic sub-states of our system, $\varrho_\ph$ and $\varrho_\el$, follow coupled equations of motion, which are given in Eq. (4) of the main text. 
The expectation values of vibrational and electronic observables, $O_\ph$ and $O_\el$, thus evolve according to the equations of motion
\begin{equation}\label{SMF}
    \frac{d}{dt}\av{O_\ph}= \frac{i}{\hbar}\av{[\tilde{H}_\ph,O_\ph]}
    \quad\text{and}\quad
    \frac{d}{dt}\av{O_\el}= \frac{i}{\hbar}\av{[\tilde{H}_\el,O_\el]} + \av{\mathcal{L}^\dagger[O_\el]},
\end{equation}
where we have used the short-hand notation $\av{\bullet}={{{\rm Tr}}}[\bullet \varrho_\ph\otimes\varrho_\el]$. 
The mean-field Hamiltonians are given by $\tilde{H}_\ph = H_\ph + \hbar W(x) s_{nn}$ and $\tilde{H}_\el = H_\el + \hbar n_1 n_2 \av{W(x)}$  with $W(x)=V_0/\hbar +\kappa_1 x + \kappa_2 x^2$ and $s_{\alpha \beta}=\av{\sigma_1^\alpha \sigma_2^\beta}$, $s_{\alpha n}=\av{\sigma_1^\alpha n_2}$, $s_{n\beta}=\av{n_1 \sigma_2^\beta}$, $s_{nn} = \av{n_1 n_2}$ with $\alpha,\beta = x, y$; the adjoint Lindblad operator reads $\mathcal{L}^\dagger[\bullet]= \gamma \sum_{k=1}^2(\sigma_k^+\bullet\sigma_k^- - \frac{1}{2}\{n_k,\bullet\})$. 
Further details can be found in the main text. 

Using Eq.~\eqref{SMF}, we find the mean-field equations of motion 
\begin{equation}
    \begin{aligned}
   \dot{\left \langle x \right \rangle} &= \omega_\mathrm{rel} \left \langle p \right \rangle,\\
    \dot{\left \langle p \right \rangle} &= - (\omega_\mathrm{rel} + \kappa_{2} s_{nn})\left \langle x \right \rangle + \kappa_{1} s_{nn},\\
    \dot{\left \langle x^2 \right \rangle} &= \omega_\mathrm{rel}(\left \langle xp \right \rangle + \left \langle px \right \rangle) \\
    \dot{\left \langle p^2 \right \rangle} &= -(\omega_\mathrm{rel} + 2\kappa_2 s_{nn})(\left \langle xp \right \rangle + \left \langle px \right \rangle) - 2\kappa_1p s_{nn} \\
    \dot{\left \langle xp \right \rangle} &= \omega_\mathrm{rel}(p^2 - x^2) - \kappa_1 x s_{nn} - 2\kappa_2 x^2 s_{nn},
    \end{aligned}
\end{equation}
for the relevant vibrational variables. 
The equations of motion for the single-body electronic variables read 
\begin{equation}
    \begin{aligned}    
    \dot{\left \langle \sigma_1^x \right \rangle} &= - \av{W(x)} s_{yn}- \Delta \left \langle \sigma_1^y \right \rangle -\frac{1}{2}\gamma \left \langle \sigma_1^x \right \rangle,\\
    \dot{\left \langle \sigma_2^x \right \rangle} &=-\av{W(x)}s_{ny} -\Delta \left \langle \sigma_2^y \right \rangle -\frac{1}{2}\gamma \left \langle \sigma_2^x \right \rangle,\\
    \dot{\left \langle \sigma_1^y \right \rangle} &= \av{W(x)} s_{xn} - 2\Omega (1 - 2\left \langle n_1 \right \rangle) +\Delta \left \langle \sigma_1^x \right \rangle-\frac{1}{2}\gamma \left \langle \sigma_1^y \right \rangle,\\
    \dot{\left \langle \sigma_2^y \right \rangle} &= \av{W(x)} s_{nx} - 2\Omega (1 - 2\left \langle n_2 \right \rangle) +\Delta \left \langle \sigma_2^x \right \rangle-\frac{1}{2}\gamma \left \langle \sigma_2^y \right \rangle,\\
    \dot{\av{n_1}} & = \Omega \left \langle \sigma_1^y \right \rangle - \gamma \left \langle n_1 \right \rangle, \\
    \dot{\av{n_2}} & = \Omega \left \langle \sigma_2^y \right \rangle - \gamma \left \langle n_2 \right \rangle
    \end{aligned}
\end{equation}
and those for the two-body variables are given by 
\begin{equation}
    \begin{aligned}
        \dot{s}_{xx} &= - \frac{1}{2}\av{W(x)}(s_{xy} + s_{yx}) - 2\Delta(s_{xn} + s_{nx}) - \Delta (\av{\sigma_1^x} + \av{\sigma_2^x}) - \gamma s_{xx}\\
        \dot{s}_{xy} &= \frac{1}{2} \av{W(x)}(s_{xx} - s_{yy}) - 4\Omega s_{xn} + 2\Omega \av{\sigma_1^x} + \Delta(s_{xx} - s_{yy}) - \gamma s_{xy}\\
        \dot{s}_{xn} &= - \av{W(x)} s_{yn} + \Omega s_{xn} - \Delta s_{yn} - \frac{3}{2} \gamma s_{xn}\\
        \dot{s}_{yx} &= \frac{1}{2} \av{W(x)}(s_{xx} - s_{yy}) - 4\Omega s_{nx} + 2\Omega \av{\sigma_2^x} + \Delta(s_{xx} - s_{yy}) - \gamma s_{yx}\\
        \dot{s}_{yy} &= \frac{1}{2} \av{W(x)} (s_{xy} + s_{yx}) - 4\Omega (s_{yn} + s_{ny}) + 2\Omega(\av{\sigma_1^y} + \av{\sigma_2^y}) + 2\Delta(s_{xn} + s_{nx}) - \Delta(\av{\sigma_1^x} + \av{\sigma_2^x}) - \gamma s_{yy}\\
        \dot{s}_{yn} &= \av{W(x)} s_{xn} + \Omega (s_{yy}- 4s_{nn}) + \Delta s_{xn} - \frac{3}{2} \gamma s_{yn}\\
        \dot{s}_{nx} &= - \av{W(x)} s_{ny} + \Omega s_{nx} - \Delta s_{ny} - \frac{3}{2} \gamma s_{nx}\\
        \dot{s}_{ny} &= \av{W(x)} s_{nx} + \Omega (s_{yy}- 4s_{nn}) + \Delta s_{nx} - \frac{3}{2} \gamma s_{ny}\\
        \dot{s}_{nn} &= \Omega(s_{yn} + s_{ny}) - \gamma s_{nn}. \\
    \end{aligned}
\end{equation}

\begin{figure}
    \centering
    \includegraphics[scale = 0.45]{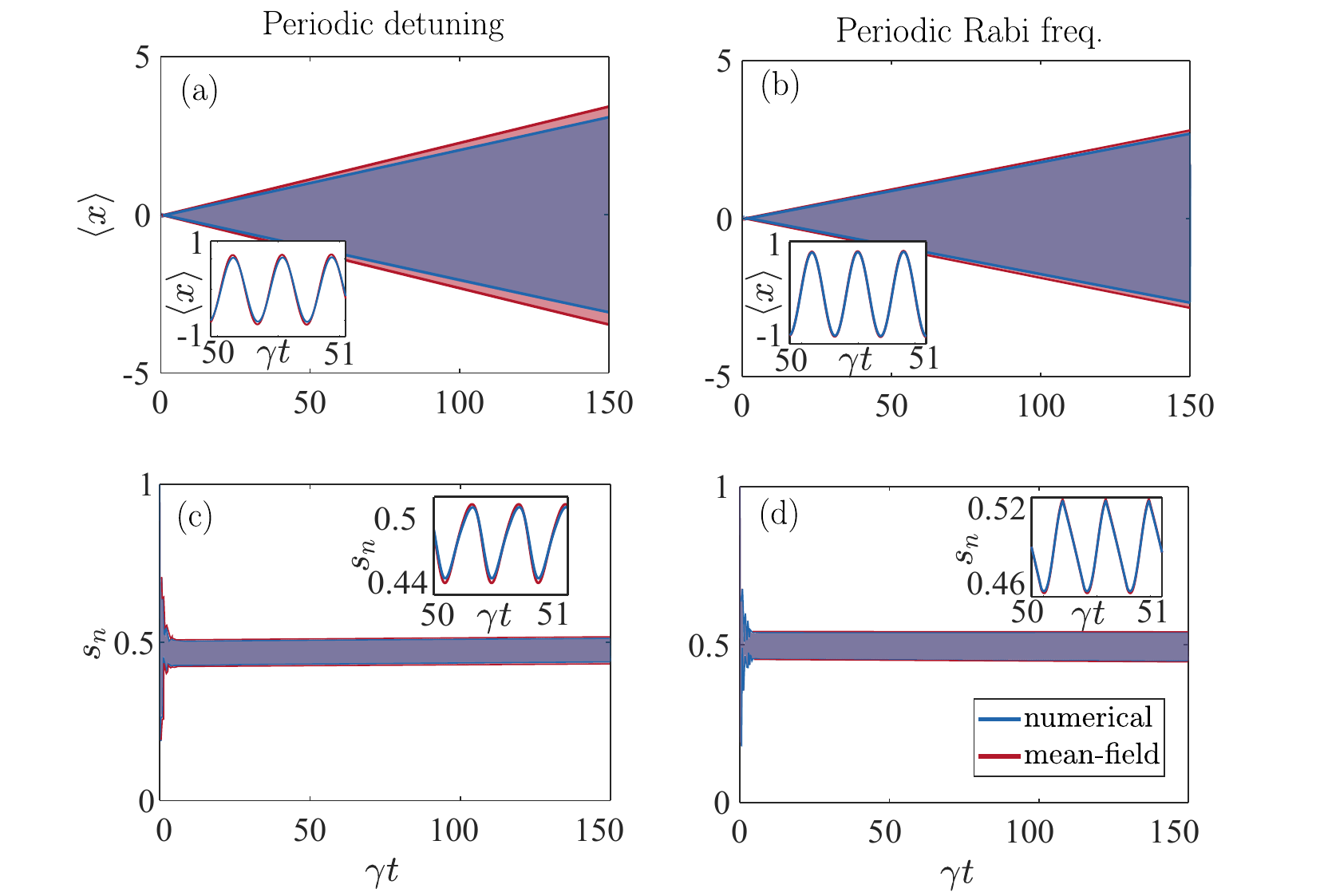}
    \caption{\textbf{Mean-field dynamics vs numerical solution.} 
    The plots (a) and (b) show the time evolution of the mean relative position of the ions $\av{x}$ as a function of the dimensionless time $\gamma t$ for periodically changing detuning $\Delta$ and Rabi frequency $\Omega$, respectively; 
    panels (c) and (d) show the corresponding evolution of the mean occupation of the double excited state $s_{nn}$. 
    Red lines indicate the solution of the mean-field equations of motion and blue lines are the results of our numerically exact calculations. 
    For all plots, we have used the same parameters as in Fig. 2 of the main text.}
    \label{fig:my_label}
\end{figure}

The mean-field equations given above form a closed set of non-linear first-order differential equations, which can be easily solved numerically. 
To confirm the validity of the mean-field approximation, we also solve the exact master equation for the full state $\varrho$ of the system, see Eq. (3) of the main text. 
To this end, we use the QuTiP package \cite{joh_2012, joh_na_no_2013} after truncating the number of vibrational modes at $N=60$. 
As shown in Fig.~\ref{fig:my_label}, the results of both approaches are in excellent agreement for the parameter regime discussed in the main text. 

In general, one would expect heating and cooling of the bosonic modes in the current system to become relevant after a certain time-scale. The presence of these effects would modify Eq. (3) as follows:
\begin{equation}
\dot{\varrho} = -\frac{i}{h}[H, \varrho] + \mathcal{L}[\varrho] + \mathcal{L}_\mathrm{C}[\varrho] + \mathcal{L}_\mathrm{H}[\varrho],
\end{equation}
where $\mathcal{L}_\mathrm{C}[\bullet] = \gamma_\mathrm{C} (a\bullet a^\dagger - \frac{1}{2}{a^\dagger a, \bullet})$ represents the dissipator associated with cooling, and $\mathcal{L}_\mathrm{H}[\bullet] = \gamma_\mathrm{H} (a^\dagger \bullet a - \frac{1}{2}{a^\dagger a, \bullet})$ the one associated with heating. 
By defining $\zeta = (\gamma_\mathrm{H} - \gamma_\mathrm{C})/2$, Eq. (5) becomes 
\begin{equation}
\langle \ddot{x} \rangle + \zeta \langle \dot{x} \rangle + \left[\orel(\orel + \kappa_2 s_{nn}) + \zeta^2\right]\langle x \rangle = -\kappa_1 \orel s_{nn}\, ,
\end{equation}
showing that these effects introduce a damping term and a change in the resonance frequency. Moreover, heating and cooling also affect the dynamics of the average excitation number $n_\mathrm{ph}$. Specifically, we have that $\dot{n}_\mathrm{ph} \rightarrow \dot{n}_\mathrm{ph} + \zeta \langle x^2\rangle + \zeta \langle p^2\rangle - \zeta + \gamma_\mathrm{H}$. The rates $\gamma_{\rm H},\gamma_{\rm C}$ for heating and cooling strongly depend on the experimental setup. However, typical values are of the order of a few kHz \cite{ra_es_2000, dan_na_2011, brow_ku_2015}.
For the parameters explored in our case study, this results in $\gamma_{\rm H},\gamma_{\rm C}/\gamma \approx 0.1, 0.01$. We can thus neglect these effects in our derivation and simply consider that they will set a maximal timescale within which the charging of the flywheel can be described as we do in the main text. 

\section{Charging limit and time of operation}

\begin{figure}
    \centering
    \includegraphics[scale = 0.42]{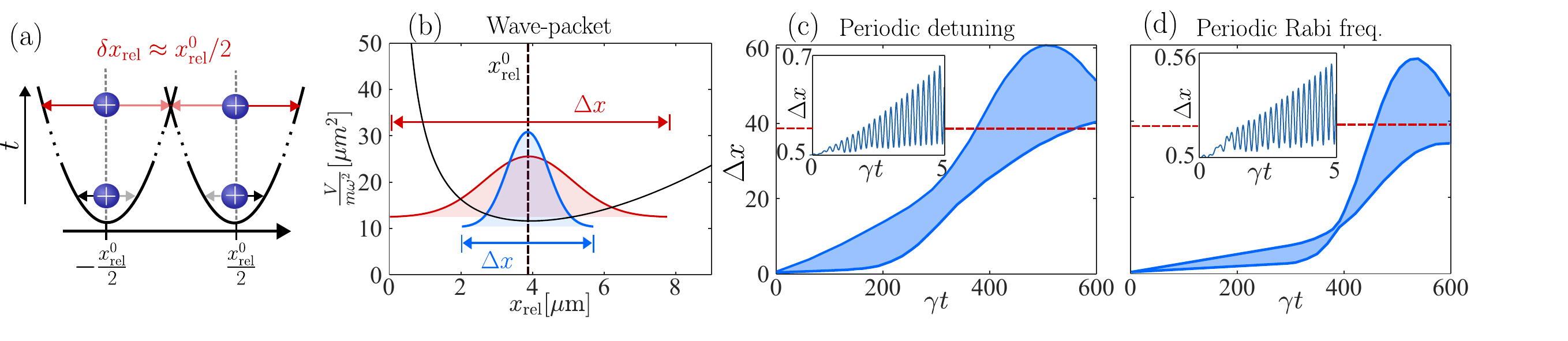}
    \vspace{-0.2cm}
    \caption{\textbf{Charging limit.} (a) The red arrows indicate the growing amplitude of the harmonic oscillations of the ions around their equilibrium positions $x^0_{1,2}=\pm \xrel^0$. 
    The charging limit is reached when the ions come close to the singularity of the Coulomb interaction potential, i.e., when their respective oscillation amplitudes reach $\xrel^0/2$. 
    (b) Spreading wave package describing the vibrational degree of freedom in the effective potential $V_\rel$. 
    The red wave package indicates the charging limit, which is reached when the width of the wave package exceeds the characteristic length scale of the potential. 
    Plots (c) and (d) show the variance $\Delta x$ as a function of the dimensionless time $\gamma t$ (blue). 
    The red dashed lines indicate the charging limit $\xrel^0/\ell_0 \approx 40$. 
    Both plots were obtained with the same parameter values as Fig. 2 of the main text and from the numerically exact solution of the master equation describing the system, where the vibrational Hilbert space was truncated at $N = 60$.}
    \label{wp_size}
\end{figure}

The charging time of our flywheel is limited by the range of validity of the approximations our model relies on. 
First, the relative oscillation amplitude has to remain small enough to ensure that the ions, which oscillate around their equilibrium positions $x^0_{1,2}=\pm \xrel/2$, do not cross the singularity of the Coulomb interaction potential, see Fig.~\ref{wp_size}(a). 
Hence, we have to impose the condition  $\ell_0 \left \langle x\right \rangle \ll \xrel^0/2$. 
Using the parameter values of the main text, this condition yields a maximal oscillation amplitude of $\left \langle x \right \rangle_\mathrm{limit} \approx 20$.     
Second, to ensure the validity of the Taylor expansion of the potential $V_\rel$, which was introduced in Eq.~\eqref{SPot}, the wave package describing the vibrational degree of freedom has to remain localized in the vicinity of the minimum of this potential. 
To meet this condition, we have to require that $\Delta x = [\av{x^2} -\av{x}^2]^\frac{1}{2}\ll \xrel^0/\ell_0$, see Fig.~\ref{wp_size}(b).
Upon calculating the variance $\Delta x$ from the numerically exact solution of the Master Equation (3) in the main text, we find that, for the parameter values used in the main text, $\Delta x$ reaches $\xrel^0/\ell_0 \approx 40$ at $t_{{{\rm limit}}}\approx 400/ \gamma $, see Figs.~\ref{wp_size}(c) and (d). 
At this time, the amplitude of $\av{x}$ is approximately $10$, which is still a factor $2$ smaller than the limit $\av{x}_{{{\rm limit}}}\approx 20$ imposed by the first condition. 
Hence, our model should be applicable for charging times $t< t_{{{\rm limit}}}\approx 400 / \gamma $.

\bibliography{supp}